\documentclass{revtex4}
\usepackage{graphicx}
\usepackage[]{epsfig}

\newcommand{\beq}{\begin{equation}}
\newcommand{\eeq}{\end{equation}}
\newcommand{\beqd}{\begin{displaymath}}
\newcommand{\eeqd}{\end{displaymath}}
\newcommand{\beqa}{\begin{eqnarray}}
\newcommand{\eeqa}{\end{eqnarray}}

\newcommand{\comment}[1]{}

\newcommand{\Tr}{{\rm Tr}\,}
\newcommand{\W}{{\mathcal W}}

\newcommand{\Q}{\tilde{Q}}
\newcommand{\br}{\mathbf{r}}

\begin{document}

\title{On Critical Dynamics in Glassy Systems}

\author{Giorgio Parisi$^{1,2,3}$ and Tommaso Rizzo$^{1,2}$} \affiliation{$^1$ Dip. Fisica,
Universit\`a "Sapienza", Piazzale A. Moro 2, I-00185, Rome, Italy \\
$^2$ IPCF-CNR, UOS Rome, Universit\`a "Sapienza", PIazzale A. Moro 2,
I-00185, Rome, Italy \\ $^3$ INFN, Piazzale A. Moro 2, 00185, Rome,
Italy}

\begin{abstract}
Critical dynamics in various glass models including those described by mode coupling theory is  described  by scale-invariant dynamical equations with a single non-universal quantity, {\it i.e.} the so-called parameter exponent that  determines  all the  dynamical  critical  exponents. We   show that these  equations follow  from the  structure of the  static  replicated  Gibbs  free  energy  near  the critical  point.  In  particular the exponent parameter is given  by the ratio between two cubic proper  vertexes that can  be expressed  as  six-point cumulants  measured in a purely static  framework.
\end{abstract}

\maketitle

\section{Introduction}

In this paper we will show that there is a deep connection between statics and dynamics in various glass models including those described by Mode-Coupling Theory. More precisely we will derive  under very general assumptions two equations (namely eq. (\ref{result}) and eq. (\ref{ris2}) below) that connect 
 the so-called dynamical parameter exponent $\lambda$ with quantities that can be either computed or measured in a purely {\it static} framework. The first equation shows that the parameter $\lambda$ is equal to the ratio $w_2/w_1$ of two cubic coefficients in the replicated Gibbs free energy and will be proven solving in parallel the statics (using replicas) and the dynamics (using the super-field formulation of Langevin dynamics). Then the ratio $w_2/w_1$ will be further identified with the ratio  $\omega_2/\omega_1$ of two cubic cumulants of the two-points order-parameter that can be measured statically. 
The two equations can be used to determine the parameter $\lambda$ without solving explicitly the dynamics of the system.  In a set of recent publication this method has been applied (without justification) to a number of mean-field spin-glass models yielding new analytical predictions \cite{calta2,ferra1,calta3}. In some cases the dynamical exponent was already known from the explicit solution of the dynamics and the novel computations in \cite{calta2,ferra1,calta3} offered an {\it a posteriori} validation of equation (\ref{result}), while in this paper we will present an {\it a priori} model-independent justification. In some other cases instead the method  yielded novel prediction for the parameter exponents that was compared with existing numerical work. Another recent application of eq. (\ref{result}) in the context of super-cooled liquids has been presented in \cite{franz12} while in \cite{Rizzo12} the same equation has been applied to Szamel's replicated Ornstein-Zernicke equations \cite{Szamel} showing that the replica method yields a characterization of critical dynamics in the $\beta$-regime of super-cooled liquids that is equivalent to the one of Mode-Coupling-Theory, both qualitatively and quantitatively.

Spin-Glass (SG) dynamics in both full Replica-Symmetry-Breaking ($f$-RSB) and one-step RSB  (1RSB) models exihibits critical slowing down \cite{Sompolinsky81,Sompolinsky82,Kirkpatrick87c,Crisanti93}. In particular it is well known that the dynamics of 1RSB models  at the dynamical transition temperature $T_d$ obeys the same dynamical equation of the schematic mode-coupling-theory (MCT) developed in the context of supercooled liquids \cite{Bengtzelius84,Gotze84,Gotze89, Gotze09,Bouchaud96}. 
In the spin-glass context one considers the relaxation of $C(t)$ {\it i.e.} the overlap between a given initial equilibrium configuration of the dynamics and the configuration at time $t$. In the discontinuous 1RSB case at $T_d$ the relaxation is a two-step process in which the system spends an increasing amount of time around a plateau value.
Models with a discontinuous 1RSB transition include the $p$-SG either spherical or Ising, the Potts SG model and the Random Orthogonal Model (ROM).
 MCT predicts that there are two exponents controlling the dynamics around the plateau, in the early $\beta$ regime $C(t)$ approaches the plateau value with a power law $C(t)\approx q_{EA}+c_a/t^a $ while in the late $\beta$ (or early $\alpha$) regime $C(t)=q_{EA}-c_b t^b$ \cite{Gotze09}. 
A well-known prediction of MCT is the following relationship between the decay exponents:
\beq
{\Gamma^2(1-a) \over \Gamma(1-2 a)}={\Gamma^2(1+b) \over \Gamma(1+2 b)}=\lambda
\label{uno}
\eeq
Where $\lambda$ is the so-called exponent parameter which is the object of this work.
In case of a continuous transition, there is no dynamic arrest
and no $b$ exponent is defined. Well known instances are, e.g., the
paramagnet to full-RSB SG transition along the de Almeida Thouless (dAT)
line in mean-field (MF) SG models, either fully connected
\cite{Sompolinsky81,Sompolinsky82} or on random graphs
\cite{Carlson90}, as well as the SG transition in Potts models with $p
\leq 4$, (both fully connected \cite{Gross85} and on the Bethe lattice
of any connectivity \cite{GOLD}), and in the $p$-spin spherical
model with large external magnetic field \cite{Crisanti93}.

We will show that the parameter exponent can be associated to physical observables computed in a static framework. We recall that in general the analytic treatment of the dynamics is more complicated than the statics and only a few models have been studied
so far: the soft-spin Sherrington-Kirkpatrick (SK) model
\cite{Sompolinsky81,Sompolinsky82}, schematic MCT's \cite{Gotze84},
soft-spin $p$-spin and Potts glass models for which notably the connection
with MCT was first identified \cite{Kirkpatrick87c}. This prompted to
consider the spherical $p$-spin SG \cite{Crisanti92,Crisanti93} in all
details as a MF structural glass, cf., e.g., Ref. \cite{Cavagna05},
even in the off-equilibrium regime below $T_d$ \cite{Cugliandolo93}.
In these cases dynamics is explicitly solved and $\lambda$
 computed exactly. In particular, one finds that {\it it is not
  universal} and depends on the model and on the external parameters.
On the other hand, its computation becomes difficult when we consider
more complicated MF systems and notably finite-dimensional ones.

It is well known \cite{Monasson95,Franz98b, Franz11b} that, similarly to the static transition, also the dynamic one can be located as the critical point of an appropriate potential that can be computed in a purely static framework.
The properties of the free energy (for the continuous transition) of the potential (for the discontinuous transition) can be described in the replica framework by a replicated Gibbs free energy
$G$. 
\begin{eqnarray}
G(\delta Q)&=&{1 \over 2}\sum_{(ab),(cd)}\delta Q_{ab} M_{ab,cd} \delta Q_{cd}
\label{f:Gibbs_fe}
\\
\nonumber
&&-{w_1 \over 6}\Tr \delta Q^3-{w_2 \over 6}\sum_{ab}\delta Q_{ab}^3
\end{eqnarray}
with $a,b,c,d=1, \ldots, n$. The order parameter $\delta Q_{ab}$ is the deviation of the two-point function relevant for the given problem, {\it e.g.} the spin overlap in spin-glass models or the density-density fluctuations in supercooled liquids. The case $n=0$ is relevant for the continuous
transition \cite{Edwards75,Sherrington75,Parisi79}, the case $n=1$ for
the dynamic discontinuous transition
\cite{Monasson95,Franz98b,Crisanti08,Franz11b}. 
The first result that we will derive in this paper is that, both in the continuous and discontinuous case, the exponent parameter $\lambda$ is given by the ratio between the coefficients $w_1$ and $w_2$ of the static replicated Gibbs free energy:
\beq
\lambda={w_2 \over w_1}
\label{result}
\eeq
We note that this ratio also yields the breaking point $x$ in the
case of continuous transition from RS to full-RSB \cite{Gross85}. 

As we already said this result has been recently confirmed {\it a posteriori} considering fully-connected models where the dynamics and the statics can be both obtained, in particular in the case of the de Almeida-Thouless line of the Sherrington-Kirkpatrick model and in the case of the spherical $p$-spin model \cite{calta1,ferra1}.
Most importantly it can be used to obtain the dynamical exponents without solving explicitly the dynamics. In particular it has been applied to various  models that were studied numerically in the past, notably the Potts SG \cite{Brangian02} and the Random-Orthogonal-Model SG \cite{Sarlat09}. 
Exploiting the fully-connected nature of these model the replicated Gibbs free energy was computed analytically by means of the saddle-point method and the previous numerical estimates were shown to be in fair agreement with the new analytical predictions \cite{calta2,calta3}.  

In the following we will argue that the above result holds provided the Gibbs free energy admits the expansion (\ref{f:Gibbs_fe}) near the critical point, including the case of finite-dimensional models above their upper critical dimension. In general however it is not possible to obtain an exact analytical expression of the Gibbs free energy. This problem is present not only in the case of finite-dimensional models but also for mean-field models defined on finite connectivity random lattices. Further progress can be made recalling that in general the Gibbs free energy is obtained as the Legendre transform of the free energy and therefore its (proper) vertexes can be associated to cumulants of the order parameter.
This simple observation leads to the second and more fundamental result that we are going to derive in this paper: the parameter exponent is given by the ratio 
\beq
\lambda={\omega_2 \over \omega_1}
\label{ris2}
\eeq
where $\omega_2$ and $\omega_1$ are six-point static functions of the microscopic variables $s_i$ (that is spins in spin-glass models or density fluctuations in structural glass models) defined by:
\beq
\omega_1 \equiv {1 \over N} \sum_{ijk}\overline{\langle s_i s_j\rangle_c \langle s_j s_k\rangle_c \langle s_k s_i\rangle_c}
\eeq
\beq
\omega_2 \equiv {1 \over 2 N} \sum_{ijk}\overline{\langle s_i s_j s_k\rangle_c^2 }
\eeq
where the overline means average over the disorder and the square bracket mean thermal average.
In the case of the discontinuous transition the above static averages  are only defined in the glassy phase because of metastability. Most importantly the glassy phase is characterized by the fact that there is an exponential number of metastable states and the above expression have to be interpreted in a slightly different way: thermal averages have to be performed inside the same metastable state, while the overline stands for both the summation over all possible states and the standard disorder average.
The above definitions therefore can be naturally extended to models with no quenched disorder (notably structural glasses), in this case the overline means just a summation over the different metastable states. 
It is well known that the MCT singularity is characterized by the divergence of the $\chi_4$, a four point susceptibility, our result shows that the parameter exponent is related to the ratio of two $\chi_6$ susceptibilities.

We devised a method to compute  analytically the above six-point functions for models  defined on finite connectivity random  lattices obtaining a prediction for the ratio $w_1/w_2$ that compares very well with the numerical simulations of the Bethe lattice SG \cite{calta1,PRR}. Besides the above cumulants can be computed numerically. As we will discuss in the paper $\omega_1$ and $\omega_2$ both diverge at criticality and this may cause huge finite size effect.

The paper is organized as follows. In section \ref{FULLY} we discuss a simple fully-connected spin-glass model where one can obtain closed equations both for the statics and the dynamics.
In spite of its simplicity the study of this model will not only confirm that (\ref{result}) holds but will also help us understand why it is so. The key ingredient is that we have to study critical dynamics at large times in an expansion around the so-called Fast Motion limit corresponding to the statics.
In section \ref{SD} we will argue that the result (\ref{result}) holds also in more general cases, in particular  we will consider three different types of SG transitions (specified by the form of the Gibbs free energy) and the corresponding critical dynamical behavior.
In section \ref{general} we will discuss the second result (\ref{ris2}) paying particular attention to the problems connected with the divergence of the physical observables $\omega_1$ and $\omega_2$. 
Technical details will be largely postponed to the appendices.
In section \ref{conclusions} we will give our conclusions. In particular we will give discuss extensively the connection between our results and previous results obtained within MCT.

\section{Fully connected Models}
\label{FULLY}

\subsection{The spherical Model}

In the case of the fully connected spherical $p$-spin models one can obtain closed equations for the dynamics.
The model is given by a set of $N$ continuous spins with the spherical constraint $\sum_i s_i^2=N$ and interacting through an hamiltonian of the form:
\beq
H=-\sum_{i_1< \dots <i_p}J_{i_1 \dots i_p}s_{i_1}\dots s_{i_p}
\eeq
where the $J$ are quenched random couplings with zero mean and variance $\overline{J^2}=p!/(2N^{p-1})$.
The order parameter is $C(t)$ defined as  the overlap  between a given equilibrium configuration at time $t=0$ and the configuration reached by the dynamics  at time $t$ averaged over the thermal noise and the disorder.
In the high temperature paramagnetic phase $C(t)$ decays to zero at large times and obeys the following equation \cite{Crisanti93,Cavagna05}:
\beq
\dot{C}(t)=-T C(t)-{p \over 2 T}\int_0^t C^{p-1}(t-u)\dot{C}(u)du
\eeq
with $C(0)=1$.
We will consider the case $p=2$ corresponding to a continuous transition.
In this case it is convenient to rewrite the above equation in the form:
\beq
T\dot{C}(t)=2\tau  C(t)-C^2(t) -\int_0^t (C(t-u)-C(t))\dot{C}(u)du\, \ \ ;
\ \tau \equiv {1-T^2\over 2}
\eeq
note that $\tau<0$ in the high temperature region where the above equation is valid.
If the last term in the r.h.s. were absent we could easily solve the
dynamical equation we would have an exponential decay at any finite
$\tau$ and a power law decay $C(t)\propto 1/t$ at $\tau=0$. The
presence of the last term instead is peculiar feature of SG dynamics
and changes completely the dynamical behaviour. One can check that  it produces a
term proportional to $\dot{C}(t)$ the dominates over the one on the
l.h.s., therefore in order to study the large times critical behaviour
at small $\tau$ we can set it to zero. We can rescale the correlation as
$C=-\tau f$ and obtain an equation that is a special case
($\lambda=0$) of the general equation \cite{Crisanti93,SommersFischer,Gotze84}:
\beq
0\,=\,2  f(x)\,+f^2(x)(1-\lambda) +\int_0^x (f(x-y)-f(x))\dot{f}(y)dy
\label{TI-SK}
\eeq
The solution of the resulting equation is invariant under a rescaling
of time $x \rightarrow b  x$, it diverges as $1/x^a$ for small $x$ and
decays exponentially at large $x$. The exponent $a$ can be determined
plugging the asynptotic form  $1/x^a$ into the equation, this yields:
\beq
\lambda={\Gamma^2(1-a)\over\Gamma(1-2a)}\ ,
\label{lambdaeq}
\eeq
therefore in the present case $\lambda=0$ we have $a=1/2$.

In order to determine the actual time scale we should match the large time
solution with the small time solution. In general we are only
interested in the divergence of the time scale as $\tau$ goes to zero
and this can be obtained considering that  the solution must not depend on
$\tau$ at any finite time, therefore we obtain:
\beq
C(t)=\tau \, f(t/t^*)\ \ \ t \ll 1,\ \   t^* \propto {1 \over \tau^{1
    \over a}}
\label{match}
\eeq
Note that in the present case $a=1/2$ and  $t^* \propto \tau^2$ which is a
completely different result from the case in which the last term in the
r.h.s. is absent and where we would have $t^* \propto \tau$.

\subsection{Soft Spin Sherrington-Kirkpatrick Model}
In this  section we will consider a more general fully connected
spin-glass model. we will show that while  it is not possible to
obtain the dynamical equations in a closed simple form as in the
spherical case, the dynamics
{\it on large time scales} reduces to the above scale invariant
equation and shares the same critical properties. In spite of its simplicity the derivation in this model contains all the ingredients that can be easily generalized to the more complex transitions considered in the following sections. 

Our result eq. (\ref{result}) connects dynamics on large time-scale with statics therefore in order to derive it, it is convenient to work in a framework where the similarity between statics and dynamics is apparent. The dynamical super-field approach has this property because it is based on a dynamical action similar to that of the static. A general presentation of the approach can be found in \cite{Zinn}. Its application to the SG problem was done by Kurchan in \cite{Kurchan92} and his results will be our starting point in the derivation.

The advantage of this formalism is that it makes apparent the equivalence between dynamics and the statics replica method.
We consider a  soft spin version of the SK model
with energy:
\beq
H=-\sum_{1 \leq i \leq j \leq N}J_{ij}s_i s_j+\sum_{i=1}^N H_0(s_i)
\eeq 
where $J_{ij}$ are quenched random couplings with zero mean and
variance $1/N$ while  $H_0(s_i)$ is a spin length probability that dominates over the
quadratic interaction for large values of $s_i$.

\subsubsection{Statics by The Replica Method}

The static of the problem can be solved by means of replicas and yields the equations:
\beq
Q_{ab}=\langle s_a s_b\rangle
\label{stateq}
\eeq 
where the square brackets means that we sum over the spins with the following weight:
\beq
\exp[{\beta^2 \over 2 }Q_{ab}s_a s_b-\beta \sum_a H_0(s_a)]
\label{statpot}
\eeq 
In the paramagnetic phase the solution is simply
\beq
Q_{ab}=q_d\delta_{ab}
\label{DIAGO}
\eeq
where  $q_d$ ($d$ stands for diagonal)  obeys the equation:
\beq
 q_d={\int  s^2 e^{{\beta^2 \over 2
    }q_d s^2- \beta H_0(s)}\, ds \over \int   e^{{\beta^2 \over 2
    }q_ds^2- \beta H_0(s) }\, ds}
\label{EQQD}
\eeq
In order to study the continuous spin-glass transition 
eq. (\ref{stateq}) is expanded in powers of 
\beq
\delta Q_{ab} \equiv Q_{ab}-q_d \delta_{ab}
\label{REPEXP}
\eeq
The coefficients of the expansion are written in terms of
spin-averages computed in the solution with $\delta Q_{ab}=0$. At the
first order for instance we need four spin averages that reads:
\beq
\langle s_a s_b s_c s_d\rangle=(q_4-3 q_d^2)\delta_{abcd}+q_d^2(\delta_{ab}\delta_{cd}+\delta_{ac}\delta_{bd}+\delta_{ad}\delta_{cb})
\label{FPF}
\eeq
where 
\beq
 q_4={\int  s^4 e^{{\beta^2 \over 4
    }q_d s^2- \beta H_0(s)}\, ds \over \int   e^{{\beta^2 \over 4
    }q_ds^2- \beta H_0(s) }\, ds}
\eeq
To go to the second order in $\delta Q_{ab}$ we need correlations
functions between six replicas, the final result is:
\beq
0 = 2 \tau \delta Q_{ab}+w_1 (\delta Q)^2_{ab}
\label{statexp}
\eeq
where
\beq
\tau \equiv {\beta^2 q_d-1 \over 2}\,, \ \ \ w_1\equiv q_d^3\beta^4\,.
\eeq
that shows that at $\tau=0$ there is a continuous SG transition from a solution $\delta Q_{ab}=0$ to a solution $\delta Q_{ab}=\tau/w_1$.

\subsubsection{Large time dynamics as an expansion around the Fast Motion solution}

Within the super-field formulation \cite{Zinn} the dynamical equations of the problem are very similar to those of the static replica treatment. They were derived by Kurchan in \cite{Kurchan92} and we will brefly quote his results which are our starting point.
In the super-field approach Langevin dynamics for the model is written as an action in terms of
bosonic $(s_i(t),\hat{s}_i(t))$  and fermionic variables
$(c_i(t),\overline{c}_i(t))$  at different times and sites. These variables
are condensed in a single super-field $\phi(1)$ by means of the
introduction of two  auxiliary fermionic  coordinates
$(\theta,\overline{\theta})$:
\beq
\phi(1)\equiv
s(t_1)+\overline{c}(t_1)\theta_1+c(t_1)\overline{\theta}_1+\theta_1
\overline{\theta}_1\hat{s}(t_1)\, , \ \ \  1\equiv (t_1,\theta_1,\overline{\theta}_1)
\eeq
In terms of these coordinates we obtain the equations \cite{Kurchan92}
\beq
Q(1,2)=\langle \phi(1)\phi(2)\rangle
\label{dyneq}
\eeq
where the square brackets mean average computed with respect to the
following  action:
\beq
S=S_{KIN}+S_{POT}
\eeq
where $S_{KIN}$ is a the dynamical part of the action:
\beq
S_{KIN}\equiv {1 \over \Gamma_0}\int d\theta\, d\overline{\theta}\,dt\, \partial_\theta
\phi(\partial_{\overline{\theta}}\phi-\theta \partial_t \phi)
\eeq
where $\Gamma_0$ is the inverse of the time scale of Langevin
dynamics and changing $\Gamma_0$ amounts to a simple rescaling of  time.
 The potential part of the action reads:
\beq
S_{POT}={\beta^2 \over 2 }\int d1 d2 Q(1,2)\phi(1) \phi(2)+\beta \int d1  H_0(\phi(1))
\label{SPOT}
\eeq
The similarity with the replica action (\ref{statpot}) is evident.
Tranforming the term $\int d1 d2 Q(1,2)\phi(1) \phi(2)$ to the
original variables one immediately recovers the well known fact that 
 the dynamic is equivalent  to that of single spin with a correlated noise and a memory term \cite{Sompolinsky81,Sompolinsky82,Mezard87}. 

First of all we want to show that the dynamical treatment is
completely equivalent to the static one. As noted already by Kurchan
this can be seen considering the fast motion (FM)  limit that amounts
to consider a very fast microscopic dynamics ($\Gamma_0 \rightarrow \infty$)  or equivalently  very
large times. In this limit the system evolves so fast that the distributions at different times are
uncorrelated, correspondingly one can write:
\beq
Q(1,2)=C(0)\delta(1,2)
\eeq
where $\delta(1,2)$ is a delta function in super-space. Note the
similarity with the replica ansatz (\ref{DIAGO}).
Plugging the above ansatz in the dynamical action (\ref{SPOT})  we
obtain that the dynamics is equivalent to the  FM dynamic in a static potential  
\beq 
V(s)=C(0)s^2+H_0(s)\, 
\label{potential}
\eeq
and the dynamical equation reduces to an equation for  $C(0)$ identical
to that obeyed by $q_d$ eq. (\ref{EQQD}) consistently with the fact
that $C(0)$ is equal to the thermal average of $s^2$ at equal times,
{\it i.e.}:
\beq
C(0)=q_d
\eeq

The above simple result illustrates how  using the FM limit one can recover in a
dynamical formulation the results
of the static obtained by means of replicas.
It is easy to see that the equivalence must hold for any observable,
in particular {\it any dynamical correlation function in the FM limit must have the same form of the corresponding correlation
function in replica space}, provided one
replaces delta functions in replica space with delta functions in
superspace.
We have already written the expression for the four point
function in the replica method, see eq. (\ref{FPF}),
the corresponding dynamical  expression in FM limit can be obtained
from it:
\beq
\langle \phi(1) \phi(2) \phi(3) \phi(4)\rangle_{FM}=(q_4-3 q^2_d)\delta(1,2,3,4)+q^2_d(\delta(1,2)\delta(3,4)+\delta(1,3)\delta(2,4)+\delta(1,3)\delta(2,4))
\label{FMcorr}
\eeq

The FM limit yields the static or infinite time limit, therefore in
order to study the dynamics  on large but finite times scale we can perform an expansion around the FM
limit.
Technically this can be view as an expansion in powers of
$1/\Gamma_0$  around zero.
The order parameter is expanded around the FM solution as: 
\beq
\delta Q(1,2) \equiv Q(1,2)-C(0) \delta (1,2) 
\eeq
note that this is similar to the replica expansion (\ref{REPEXP})  in terms of the
off-diagonal order parameter.
In the FM limit we have $\delta Q(1,2)=0$ while we expect that at
finite times for a non-zero but small $1/\Gamma_0$, $\delta Q(1,2)$
will be also small. Note however that as soon as we consider a finite
albeit small value of $1/\Gamma_0$ there is always a region of small
times where $\delta Q(1,2)$ is not small and in this region it is not
appropriate to employ a perturbative expansion. 
On the other hand it can be checked self-consistently at the end that these large  small-time-differences correlations  do not contribute to the equation for the small large-times-differences  correlations. 
The behavior of the correlations at large times will turn out to have a great deal of universality while the small times behaviour is strongly model dependent.
 However because
of this scale separation of the dynamics  the solution at large times will be obtained up to an irrelevant non-universal factor that should be determined through a matching with the short-time solution. 

The expansion of the dynamical equation (\ref{dyneq}) around the FM
solution is formally identical to the expansion of the static equation
(\ref{stateq}) around the paramagnetic solution. This is a consequence
of the similarity between (\ref{SPOT}) and (\ref{statpot}). 
The expansion is
written in terms of averages with respect to a dynamics in the
potential (\ref{potential})  with the given value of $\Gamma_0$.
At leading order we have:
\beq
\delta Q(1,2)=\int d3\,d4\,\langle \phi(1)\phi(2)\phi(3)\phi(4)\rangle_{\Gamma_0}
\delta Q(3,4) +O(\delta Q^2)
\eeq
We have already seen that the in the FM limit $\Gamma_0\rightarrow
\infty$ the dynamical correlations are formally the same of the
replica treatment and the expansion of these correlations in terms of
$1/\Gamma_0$ generates terms proportional to time derivatives of $\delta
Q(1,2)$. On the other hand we have seen in the previous section that
the terms involving explicitly time derivatives play a sub-leading role
in critical SG dynamics.
Therefore at leading order we can replace the dynamical correlation at
finite $1/\Gamma_0$ with their FM limit, {\it e.g.}
eqs. (\ref{FMcorr}).
Given that FM correlations are equivalent to correlations in replica
space it follows that the expansion of the dynamical equation is
equivalent to the corresponding expansion (\ref{statexp})  in replica space, {\it i.e.}
it reads:
\beq
0 =2 \tau \delta Q(1,2)+w_1 \int d3 \delta Q(1,3)\delta Q(3,2)
\label{dyneq3}
\eeq 
with the {\it same} values of $\tau$ and $\omega_1$.
The final step of our derivation is to show that the above dynamical
equation is of the same form  (\ref{TI-SK}) obtained starting from the
spherical model.
In order to do that we recall that the order parameter
$Q(1,2)$ is a  two point superfield correlation whose most general
causal form is (see {\it e.g.} \cite{Kurchan92}):
\beq
Q(1,2)=\left\{   1+ {1 \over 2}(\overline{\theta}_1-\overline{\theta}_2)[\theta_1+\theta_2-(\theta_1-\theta_2)\epsilon(t_1-t_2){\partial
\over \partial t_1}]\right\}C(t_1-t_2)
\label{genFDT}
\eeq   
Where $\epsilon(t)$ is the sign function.
The above expression encodes causality, the
fluctuation-dissipation-theorem (FDT) and time-translational
invariance.
From the above formula we see that time-traslational invariance and FDT imply that all the components of $Q(1,2)$ can be recovered from the boson-boson component $C(t)$, in other words the whole function $Q(1,2)$ is actually described by a single real even function $C(t)$ that is the correlation $\langle s(t_1)s(t_2)\rangle$.
Note also the from (\ref{genFDT}) we see that selecting the boson-boson component in $Q(1,2)$ is formally equivalent to evaluate it for   
$\theta_1=0,\overline{\theta}_1=0,\theta_2=0,\overline{\theta}_2=0$.

Note that the object $\delta Q(1,2) \equiv Q(1,2)-\delta(1,2)C(0)$ is also of the form $(\ref{genFDT})$ with a function $\delta C(t)$ such that $\delta C (0)=0$ and $\delta C (t)=C(t)$ for $t\neq 0$. It can be also checked that the term $\int \delta Q(1,3)\delta Q(3,2)d3$ is also of the form (\ref{genFDT}) with an appropriate boson-boson component that we will express in terms of $C(t)$. These properties guarantees that if we satisfy the boson-boson component of eq. (\ref{dyneq3}) all others components will be automatically satisfied. As we said before this is equivalent to formally set $\theta_1=0,\overline{\theta}_1=0,\theta_2=0,\overline{\theta}_2=0$ and the final result is 
\beq
\left. \int \delta Q(1,3)\delta Q(3,2)  d3 \right|_{\theta_1=\theta_2=\overline{\theta}_1=\overline{\theta}_2=0}=-C^2(t)-\int_0^t (C(t-y)-C(t))\dot{C}(y) dy\ .
\label{formuletta}
\eeq 
where $t=t_1-t_2$.
The above simple result is the key ingredient of the statics-dynamics connection and it is derived in detail in   appendix \ref{appfor}.
Putting everything together we see that the dynamical equation is the
same of the spherical model obtained previously:
\beq
0=2\tau  C(t)+w_1\left(-C^2(t) -\int_0^t (C(t-u)-C(t))\dot{C}(u)du\right)\,
\label{scale-inv}
\eeq
To summarize, the results of  this subsection are:
\begin{itemize}
\item The dynamics on large time scales near the critical SG
  temperature can be obtained as an expansion around the fast motion FM
  solution. 

\item  The resulting dynamical equations  do not depend  on the terms
  that contain explicit time derivatives and therefore  are
  invariant under a rescaling of times, independently of the details
  of the model the equation reduces to eq. (\ref{scale-inv}).  

\item The coefficients of the relevant terms of the equations can be
  computed in the FM limit and thus are the same of the corresponding
  replicated static equation. The peculiar nature of SG
  critical dynamics is determined by the presence of a term $w_1  (\delta Q^2)_{ab}$
in the replica equations that comes from a term $w_1 \Tr Q^3$  in the static replicated action.

\end{itemize}

Since the dynamical equation is time scale invariant the large time
behavior is determined minus an overall time scale. This should be
determined matching the large time solution with the small time
solution which depends on the details of the dynamics and of the
model. On the other hand the divergence of the time scale approaching
the critical temperature can be obtained using matching arguments as in eq. (\ref{match}) .

\section{From Replicated Gibbs Free energy to Critical Dynamics}
\label{SD}

\subsection{The general case}

In the preceding section we have considered the case of fully connected models for which one can obtain both the static and the dynamics from the saddle point method. Then we took the FM limit in the dynamical equations and we recovered the statics. The first correction to the FM limit yielded an equation invariant under a rescaling of time which determines critical behavior, notably the dynamical exponents controlling the time decay of the correlation at criticality and the divergence of the time scale as the control parameter (temperature, field) approaches its critical value. The explicit computation of the previous section shows that there is perfect equivalence between the dynamical equations and the statical equations in the sense that their coefficients  are the same in the FM limit.
This result could have been anticipated, after all we knew from the beginning that the dynamics in the large time limit must reproduce the static. This observation allows to generalize considerably the results of the previous section and prove in full generality eqs. (\ref{result}) and (\ref{ris2}). In order to do so we will introduce the Replicated Gibbs free energy and the Dynamical Gibbs free energy and the corresponding equations of state in order to describe the statics and the dynamics of a general model.  For a generic model these quantities provide the generalization of the saddle-point equations discussed in the previous section.

We proceed in full generality considering  $n$ replicas of a a system of $N$ spins $s_i$ specified by a Hamiltonian $H_J(s)$ depending on some quenched parameter. Averages in the replicated system can be rewritten as
\beq
\langle \cdots \rangle \equiv \overline{ \langle \cdots \rangle_J}
\eeq
Where $\langle \cdots \rangle_J$ are thermal averages at fixed couplings $J$ while the overline is the average over the couplings that must be performed reweighting  each disorder realization with the single system partition function to the power $n$: 
\beq
\overline{O_J}=\frac{\int dP(J) O_J Z_J^n}{\int dP(J) Z_J^n}
\eeq
We define the following free energy functional:
\beq
F(\lambda) \equiv -{1 \over N}\ln \langle e^{\sum_{(ab)}N\lambda_{ab}\delta \Q_{ab}}\rangle
\label{FDEF1}
\eeq
where
\beq
\delta \Q_{ab} \equiv {1 \over N}\sum_i s_i^a s_i^b- q
\eeq
and $q$ is the average value of the overlap in the absence of the replicated fields $\lambda_{ab}$.
\beq
q \equiv {1 \over N}\sum_{i}\langle s_i^a s_i^b \rangle_{\lambda_{ab}=0} = \sum_i \overline{\langle s_i\rangle_J^2}
\eeq
The average value of the overlap is given by the derivative of the Free energy with respect to the fields:
\beq
\langle \delta \Q_{ab} \rangle=-{\partial F \over \partial \lambda_{ab}}
\eeq
The Replicated Gibbs free energy is defined as the Legendre transform of $F(\lambda)$:
\beq
G(\delta Q) \equiv F(\lambda)+\sum_{(ab)}\lambda_{ab}\delta Q_{ab}
\eeq
where $\lambda$ is now a function of $\delta Q_{ab}$ according to the following implicit equation:
\beq
\delta Q_{ab}=-{\partial F \over \partial \lambda_{ab}}
\eeq
Correspondingly the derivative of the Gibbs free energy yields the equation of state:
\beq
\lambda_{ab}={\partial G \over \partial \delta Q_{ab}} \ .
\eeq
In a Ginzburg-Landau sense the assumption that a given model undergoes a phase transition at some point corresponds to the assumption that the Gibbs free energy has a certain form near the critical point. 
The case of the preceding section corresponds to an equation of state of the form (setting $\lambda_{ab}=0$):
\beq
0=2 \tau \delta Q_{ab}+w_1 (\delta Q^2)_{ab}+O(Q^3)
\label{expgibb}
\eeq
We introduce a dynamical analog of the Gibbs free energy as the Legendre transform of a dynamical free energy defined as:
\beq
F^{Dyn}(\lambda) \equiv -{1 \over N}\ln \langle e^{N \, \int d1 d2 \lambda(1,2)\delta \Q(1,2)}\rangle
\label{FDEF2}
\eeq
where the square bracket mean average in the super-field variables introduced in the preceding section.
The dynamical analog of the overlap in the previous equation is the super-overlap:
\beq
\delta \Q(1,2) \equiv {1 \over N}\sum_i [\phi_i(1) \phi_i(2)- \langle \phi_i(1) \phi_i(2)\rangle^{FM}_{\lambda=0}]
\eeq
The average value of the super-overlap is given by the derivative of the dynamic Gibbs Free energy with respect to the fields:
\beq
\langle \delta \Q(1,2) \rangle=-{\partial F^{Dyn} \over \partial \lambda(1,2)}
\eeq
The dynamical Gibbs free energy is defined as the Legendre transform of $F^{Dyn}(\lambda)$:
\beq
G^{Dyn}(\delta Q) \equiv F(\lambda)+\int d1 d2 \lambda(1,2)\delta Q(1,2)
\eeq
where $\lambda$ is now a function of $\delta Q$ according to the following implicit equation:
\beq
\delta Q(1,2)=-{\partial F \over \partial \lambda(1,2)}
\eeq
Correspondingly the derivative of the dynamical Gibbs free energy yields the dynamical equation of state:
\beq
\lambda(1,2)={\partial G \over \partial \delta Q(1,2)} \ .
\eeq
Now it is well-known that the coefficients of the dynamical free energy in powers of $\lambda(1,2)$ can be expressed as cumulants of $\delta \Q$ \footnote{This result will be used in the following section, see also the detailed derivation in appendix \ref{fre}}. According to the discussion in the previous section it is clear that in the FM limit the structure of these dynamical cumulants is the same of the corresponding replicated objects. From this observation it follows that for a generic model if the replicated equation of state has a given structure, say eq. (\ref{expgibb}) above, {\it the dynamical equation of state in the FM limit has precisely the same structure with the same coefficients}:
\beq
0=2 \tau \delta Q(1,2)+w_1 \int \delta Q(1,3) \delta Q(3,2)+O(Q^3)
\eeq
In the FM limit the above equation has the trivial solution $\delta Q(1,2)=0$. Next we consider the first correction to the FM limit assuming a small but finite value of the parameter of the Langevin equation $1/\Gamma_0$.
This will produce a small modification of the parameters of the dynamical equation of state introducing additional terms proportional to the times derivatives of the correlations.
The appearance of these time derivatives will have a dramatic effect at small time differences. Indeed in the FM limit the correlation $C(t_1-t_2)$ jumps from $C(0)\neq 0$ at $t_2=t_1$ to $C(t_2-t_1)=0$ instantaneously at $t_2=t_1^{+}$ or $t_2=t_1^-$ and the presence of time derivatives will have the effect of smoothing the correlation in time.
As we have already noticed as soon as we perturb the FM limit there is always a region of time differences of order $1/\Gamma_0$ where the dynamics cannot be obtained using the above perturbative expansion in $\delta Q(1,2)$ because $\delta Q(1,2)$ is large in this region. However if we consider time differences large with respect to $1/\Gamma_0$ we can use the  equation above looking for the non-zero solution $\delta Q(1,2)$ invariant under a rescaling of time and conclude that the critical dynamics is determined by equations (\ref{TI-SK},\ref{lambdaeq},\ref{match}) with $\lambda=0$. This determines the large times behavior up to an irrelevant numerical prefactor that depends on the precise form of the correlation in the small time region. 

The key point is that all these properties can be obtained solely under the assumption that the replicated equation of state admits the expansion (\ref{expgibb}) near the critical point. This remains true also in finite dimension provided we stay above the upper critical dimension.

In the following we will consider various types of SG transitions specified by the structure of the Gibbs free energy near the critical point and we will use the static-dynamic mapping to determine the corresponding critical dynamics. The structure of the dynamical equations we will obtain is not at all new although we will offer a concise and unified derivation. The new result is that in all the cases we will consider below the exponent parameter $\lambda$ is determined by the ratio between the cubic coefficients of the static replica Gibbs free energy.
In the next section we will further argue that the cubic coefficients of the replicated Gibbs free energy can be associated to appropriate cubic cumulants of the correlations bridging the dynamical exponent with physical observables.

\subsection{Continuous Transition in zero field}
\label{contnof}
This type of transition is characterized by the following structure of the Gibbs free energy:
\beq
 G({Q}) = \frac{\tau}{2} \sum_{a,b} Q_{ab}^2 
-\frac{w_1}{6} \Tr Q^3 
- \frac{w_2}{6} \sum_{a,b}Q_{ab}^3
 \eeq 
with $q_{aa}=0$, note the presence of the term $ \sum_{a,b}Q_{ab}^3$ that was absent in
the models considered earlier and that change the coefficient of the
term $C^2(t)$ in the dynamical equations (\ref{scale-inv}). Such a structure describes  {\it e.g.}
the continuous transition in the case of the Potts model \cite{Gross85}. Another instance is given by the continuous SG transition of the SK model in presence of a small $p$-spin term with $p=3$ in both the Ising and spherical case.
In order to study critical dynamics at large time we can just extend the results of the previous sections. 
The correlation at large time-differences is described by:
\beq
C(t)=\tau \, f(t/t^*)\ \ \ t \gg 1,\ \   t^* \propto {1 \over \tau^{1
    \over a}}
\eeq
where the exponent $a$ obeys:
\beq
{w_2 \over w_1}={\Gamma^2(1-a)\over\Gamma(1-2a)}\ ,
\eeq
and the function $f(x)$ obeys the scale invariant equation:
\beq
0\,=\,  f(x)\,+f^2(x)\left(1-{w_2\over w_1}\right) +\int_0^x (f(x-y)-f(x))\dot{f}(y)dy
\eeq
the solution of the above equation diverge as $1/x^a$ for $ x
\rightarrow 0$ and goes exponentially to zero  for $x \rightarrow \infty$.

\subsection{Continuous Transition in a field}
\label{contf}
This transition is described by the vanishing of the replicon
eigenvalue in a replica symmetric Gibbs free energy with $n=0$ \cite{Temesvari02b}.
In the SK model the position of this transition in the temperature/magnetic-field plane defines the de Almeida Thouless line \cite{Mezard87}.
The replicated order parameter above the transition is replica symmetric (RS) and can be
written as:
\beq
q_{ab}^{RS}=\delta_{ab}(q_d-q_{EA})+q_{EA}
\eeq
From this expression the corresponding dynamical order parameter in the FM limit is:
\beq
Q_{FM}(1,2)=\delta(1,2)(C(0)-C(\infty))+C(\infty)\, ,\ \ \ C(0)=q_d,\ C(\infty)=q_{EA}
\eeq
Both $q_d$ and $q_{EA}$ are regular at the transition as a function of
the external parameter because the longitudinal eigenvalue is regular \cite{Temesvari02b}.
The replicated equation of the order parameter  expressed in terms of
the deviation from the RS solution $\delta q_{ab}=q_{ab}-q^{RS}_{ab}$ is:
\beq
0=r \delta q_{ab}+m_2 \left(\sum_c \delta q_{bc}+\delta q_{ac}\right)+m_3
\sum_{cd}\delta q _{cd}+ w_1 (\delta q^2)_{ab} 
- w_2 \delta q_{ab}^2
\eeq
where $r$ is the replicon eigenvalue that vanishes linearly while the relevant external parameter
(temperature, field) approaches its critical value; instead $m_2$,
$m_3$, $w_1$ and $w_2$ remain finite and as usual can be
assumed to be constant.
The corresponding dynamical equation expressed in terms of the
deviation from the FM solution  $\delta
Q(1,2)\equiv Q(1,2)-Q_{FM}(1,2)$ is :
\beq
0=r \,  \delta Q(1,2)+m_2 \, \left( \, \int d3
\, \delta Q(1,3)+\delta Q (2,3)\right)+m_3 \,
\int d3 d4 \delta Q (3,4)+ w_1 \,  \int d3 \delta Q (1,3)\delta Q(3,2) 
- w_2\, \delta Q(1,2)^2
\eeq
The above equation  is similar to that corresponding to zero
field except for  the two terms proportional to $m_2$
and $m_3$. This two terms however give a vanishing contribution
indeed:
\beq
\int d2 \delta Q(1,2)=\int d2 Q(1,2)-\int d2 Q_{FM}(1,2)=\int d2 Q(1,2)-(C(0)-C(\infty))=0
\eeq
therefore the critical behaviour of the decay of the correlation to
its infinite time limit is the same of the zero field case:
\beq
C(t)-C(\infty)=r  \, f(t/t^*)\ \ \ t \gg 1,\ \   t^* \propto {1 \over r^{1
    \over a}}
\eeq
where the exponent $a$ obeys:
\beq
{w_2 \over w_1}={\Gamma^2(1-a)\over\Gamma(1-2a)}\ ,
\eeq
and the function $f(x)$ obeys the scale invariant equation:
\beq
0\,=\,  f(x)\,+f^2(x)\left(1-{w_2\over w_1}\right) +\int_0^x (f(x-y)-f(x))\dot{f}(y)dy
\eeq
the solution of the above equation diverge as $1/x^a$ for $ x
\rightarrow 0$ and goes exponentially to zero  for $x \rightarrow \infty$.

\subsection{Discontinuous Transition}
\label{discot}

This transition can be described by a Replica Symmetric Gibbs free energy with
$n=1$ replicas (see discussion at the end of appendix \ref{fre}).
The order parameter at the critical point is given by:
\beq
q_{ab}^{RS}=\delta_{ab}(q_d-q_{EA})+q_{EA}
\eeq
The variational equation near the critical temperature is:
\beq
0=\tau + m_2 \left(\sum_c \delta q_{bc}+\delta q_{ac}\right)+m_3
\sum_{cd}\delta q _{cd}+ w_1 (\delta q^2)_{ab} 
+ w_2 \delta q_{ab}^2
\label{disa}
\eeq
Where $\tau$ vanishes linearly approaching the transition and $\delta q_{ab}$ is the difference between order parameter $q_{ab}$ and the
solution at criticality corresponding to $r=0$ and $\delta q_{ab}=0$. Note that this
definition is different from those we employed in the previous case
where $\delta q_{ab}$ was the difference between the order parameter
and solution at the given value of $r$ and therefore there was no
constant term in the equation.
At $n=1$ not only the replicon eigenvalue vanishes but also the
longitudinal one \cite{Franz11b,Temesvari02b} this is connected with the fact that
although $m_2$ is finite at the transition it gives a vanishing
contribution at $n=1$ {\it i.e.} for $\delta q_{ab}=\delta q$ we have:
\beq
0=\tau +2  m_2(n-1)\delta q+m_3 n(n-1)\delta q
+(n-2) w_1 \delta q^2 
+ w_2 \delta q^2 
\eeq
at $n=1$ the linear term disappears and the equation becomes:
\beq
 0=\tau
+ (w_1-w_2) \delta q^2 
\eeq
this reflects the fact that below $T_d$ there is a static solution
while above there is not.
In order to recover the above results in a dynamical context it is useful
to consider the dynamics starting from an equilibrated initial
configuration.
A dynamical Gibbs free energy formally identically to the replicated one can be
obtained introducing a super-field $\phi_i(1)$ where $1$ is a coordinate that
can  either select the real replica that specifies the initial condition, in this case $\phi_i(1)=s_i^{init}$  or can take value
$(t_1,\theta_1,\overline{\theta}_1)$ in this case $\phi_i(1)$ is the
standard super-field at time $t_1$. It can be checked that the
corresponding structure is equivalent to the case of $n=1$ replicas,
in particular we have:
\beq
\int d1=1
\eeq
at variance with the case treated in the previous sections
corresponding to $n=0$ where we have $\int d1=0$.
The corresponding FM dynamical solution can be written as:
\beq
Q_{FM}(1,2)=\delta(1,2)(C(0)-C_p)+C_p
\eeq
and in any given model one can check that the corresponding equations for $C(0)$ and
$C_p$ (the plateau value)  are precisely the same obtained in the replicated treatment for
respectively $q_d$ and $q_{EA}$.
In fully connected models one can follow the derivation of the previous sections obtaining that the dynamics near the dynamical temperature can be described by the same equation (\ref{disa}) of the
replicated treatment expressed in terms of:
\beq
\delta Q(1,2)=Q(1,2)-Q_{FM}(1,2)
\label{defdyn}
\eeq
The corresponding dynamical equation is therefore:
\beq
0=\tau + m_2 \int  (\delta Q(1,3)+\delta Q(2,3))d3+m_3
\int d3 d4 \delta Q(3,4)+ w_1 \int \delta Q(1,3) \delta Q(3,2)d3
+ w_2 \delta Q(1,2)^2
\label{DYNDYN}
\eeq
In order to study the above equation we choose to evaluate it with  $1$ as the initial
condition and $2 \equiv (t_2=t,\theta_2=0,\overline{\theta}_2=0)$.
We start noticing that the quadratic term can be written as:
\beqa
\int \delta Q(1,3) \delta Q(3,2)d3 &  = &  \int Q(1,3)Q(2,3)d3-2
Q(1,2)(C(0)-C_p)+
\nonumber
\\
& - & C_p
\int Q(2,3)d3-C_p\int Q(1,3)d3+2 C_p(C(0)-C_p)+C_p^2\int d3
\eeqa
computing the various terms similarly to appendix \ref{appfor} we obtain:
\beqa 
\int Q(1,3) Q(2,3)d3 & = & C(0)C(t)-\int_0^tC(t-y)\dot{C}(y)dy
\\
\int Q(1,3)d3 & = & C(0)
\label{Q13}
\\
\int Q(2,3)d3 & = & C(t)-\int_0^t\dot{C}(y)dy=C(0)
\label{Q23}
\eeqa
putting everything together we have:
\beq
\int \delta Q(1,2)\delta Q(2,3)  d2 = -\delta C^2(t)-\int_0^t (\delta
C(t-y)-\delta C(t))\delta\dot{C}(y) dy\, , \ \ \  \delta C(t) \equiv C(t)-C_p
\eeq 
the two linear terms proportional to $m_2$ and $m_3$ vanish as can be
seen from eqs. (\ref{Q13}) and (\ref{Q23}) and the the fact that $\int Q_{FM}(1,3)d3=C(0)$.
Therefore the dynamical equation reduces to the following equation for the correlation:
\beq
0=\tau -\left(1-{w_2\over w_1}\right)\delta C^2(t)-\int_0^t (\delta
C(t-y)-\delta C(t))\delta\dot{C}(y) dy
\eeq
where the inessential rescaling $\tau \rightarrow \tau w_1$ was performed.
The above equation is the same that it is obtained in schematic MCT theories \cite{Gotze84}, according to it
 the critical behaviour of the correlation around the plateau
value near the dynamical temperature $T_d$ is given by:
\beq
\delta C(t) \equiv C(t)-C_p=|\tau|^{1/2}  \, f_{\pm}(t/t^*)\ \ \ t \gg 1,\ \   t^* \propto {1 \over |\tau|^{1
    \over 2 a}}
\label{SVDYN1}
\eeq
where the function $f_-$ has to be chosen above the dynamical
temperature ($\tau<0$) while the function $f_+$ has to be choose below the
dynamical temperature ($\tau>0$).
The exponent $a$ obeys:
\beq
{w_2 \over w_1}={\Gamma^2(1-a)\over\Gamma(1-2a)}\ ,
\eeq
and the function $f_{\pm}(x)$ obeys the scale invariant equation:
\beq
\pm 1=f_{\pm}^2(x)\left(1-{w_2\over w_1}\right) +\int_0^x (f_{\pm}(x-y)-f_{\pm}(x))\dot{f}_{\pm}(y)dy
\label{SVDYN2}
\eeq
As we have said this coincides with the equation of the critical regime of schematic MCT, {\it i.e.} eq. 6.55a in \cite{Gotze84} with the exponent parameter $\lambda$ given by the dimensionless ratio $w_2/w_1$.
The solution of the above equation diverges as $1/x^a$ for $ x
\rightarrow 0$ for both $f_{+}$ and $f_{-}$. The behavior at large
value of $x$ instead is completely different.
Above the dynamical temperature we have to choose $f_-$ that goes to
$-\infty$ as $x^b$ for large $x$ where $b$ is given by the well-known
equation:
\beq
{w_2 \over w_1}={\Gamma^2(1-a)\over\Gamma(1-2a)}={\Gamma^2(1+b)\over\Gamma(1+2b)}\ .
\eeq
Below the dynamical temperature instead we have to choose $f_+$ that
decays exponentially to the constant $(1-\lambda)^{-1/2}$ for $x\rightarrow
\infty$ \footnote{We note that the critical equations of the $n=0$ case can be reduced through a constant shift to the equation for $f_+$}.
We note that the mapping between statics and dynamics can be used only for $\tau>0$ because the statics is only defined in the glassy phase. However once the dynamical equation is obtained in the glassy phase from the static one it is natural to argue the it can be continued to the liquid phase simply assuming that $\tau<0$ and that the parameter exponent  $\lambda=w_2/w_1$ is the same above and below the dynamical transition temperature.

\section{Replicated Gibbs Free energy and the physical observables}
\label{general}

In the previous section we have considered various types of SG phase transitions characterized by the form of the replicated Gibbs Free energy. We have argued that to each replicated Gibbs free energy corresponds a dynamical Gibbs free energy with the same coefficients in the FM limit. The Dynamical Gibbs free energy in the FM limit determines the nature of critical slowing down at the transition, in particular the dynamical exponents are determined by a single non-universal exponent parameter $\lambda$ which we identified with the ratio $w_2/w_1$ between the cubic coefficients.

In this section we will further explore the connection between statics and dynamics building on the fact that the Gibbs free energy is defined as the Legendre transform of the free energy and therefore the coefficients $w_1$ and $w_2$ can be expressed in terms of cumulants of the order parameter. The details of the derivation will be given in appendix \ref{fre} and \ref{inverse} while in the following we will present and discuss the result.  

The different types of SG transitions discussed in the previous section can all be associated to a replica-symmetric Gibbs free energy with either $n=0$ or $n=1$. For a general replicated spin-glass model the Replica Symmetric Gibbs free energy reads:
\beq
G(\delta Q)={1 \over 2}\sum_{(ab),(cd)}\delta Q_{ab} M_{ab,cd} \delta Q_{cd}-{w_1 \over 6}\Tr \delta Q^3-{w_2 \over 6}\sum_{ab}\delta Q_{ab}^3
\label{gibbsfre}
\eeq
As explained in appendix \ref{fre} the above expression is the Legendre transform of the free energy in presence of an appropriate field and therefore the various coefficients can be associated to spin averages. In particular $M_{ab,cd}$ is the inverse of the replica symmetric dressed propagator $G_{ab,cd}$. Due to replica symmetry the dressed propagator has three possible values depending on the number of replica indexes that are equal: $G_{ab,ab}=G_1$, $G_{ab,ac}=G_2$ and $G_{ab,cd}=G_3$. The three propagators are associated to cumulants of four spins or equivalently to two-point functions of the overlap between different replicas:
\beqa
G_1  & \equiv   N \langle \delta \Q_{12}^2\rangle = & {1 \over N}  \sum_{ij}(\overline{\langle s_i s_j\rangle^2}-q^2) 
\\
G_2 & \equiv  N \langle \delta \Q_{12} \delta \Q_{23}\rangle = & {1 \over N}\sum_{ij}(\overline{\langle s_i s_j\rangle \langle{s_i}\rangle \langle{s_j}\rangle}-q^2) 
\\
G_3 & \equiv  N \langle \delta \Q_{12} \delta \Q_{34}\rangle = & {1 \over N}  \sum_{ij}(\overline{ \langle{s_i}\rangle^2 \langle{s_j}\rangle^2}-q^2)
\eeqa
where we recall from the preceding section that 
\beq
q \equiv {1 \over N}\sum_i \overline{\langle s_i \rangle^2}
\eeq
is the Edwards-Anderson parameter and 
\beq
\delta \Q_{12} \equiv {1 \over N}\sum_i s_i^1 s_i^2-q 
\eeq
is defined as the deviation of the overlap between two replicas with respect to its average value.

As usual in the spin-glass context the overline means average with respect to the disorder while the square bracket means thermal average at fixed realization of the disorder. In a numerical simulation these objects can be computed studying the evolution of four different replicas with the same disorder but different thermal histories. The fluctuations of the various overlaps are related to $G_1$,$G_2$ and $G_3$ according to the above formulas. We stress that when the propagators are written in terms of fluctuations of the overlaps the thermal and disorder averages can be grouped in a single average that we also represent with an angle bracket with a slight abuse of notation. 
The above formulas are valid in the case $n=0$, in the case $n=1$ they have to be interpreted in a slightly different way. This point is discussed more extensively at the end of appendix \ref{fre}, here we recall that the case $n=1$ describes a glassy phase characterized by the fact that there is an exponential number of metastable states. In this case we have to use the prescription that (see e.g. \cite{Franz11b}) {\it thermal averages in the r.h.s. of the above expressions has to be performed {\it inside} the same metastable state and the overline must be interpreted as an average over both the disorder and the many states at given disorder}. In a numerical simulation one should then consider four replicas of the system with different thermal histories but with the {\it same} equilibrium initial condition. Indeed the initial condition selects a given metastable state and different initial conditions correspond to different metastable states. We see that this framework can be easily extended to systems (notably structural glass models) with no quenched disorder but with self-induced disorder caused by the splitting of the equilibrium state into many amorphous components, in this case the overline in the above expression means just average with respect to the different components.

The cubic coefficients of the Gibbs free energy turn out to be given by (see appendix \ref{inverse}):
\beq
w_1=r^3 \omega_1 \ \ \ w_2=r^3 \omega_2
\label{ow}
\eeq
where $r$ is the replicon {\it i.e.} the inverse of the spin glass susceptibility that diverge at criticality: 
\beq
r \equiv {1 \over G_1-2 G_2+G_3}= \left[ {1 \over N}\sum_{ij}\overline{\langle s_i s_j\rangle_c^2} \right]^{-1}
\eeq
and $\omega_1$, $\omega_2$ are six-point functions given by:
\beq
\omega_1={1 \over N} \sum_{ijk}\overline{\langle s_i s_j\rangle_c \langle s_j s_k\rangle_c \langle s_k s_i\rangle_c}
\eeq
\beq
\omega_2={1 \over 2 N} \sum_{ijk}\overline{\langle s_i s_j s_k\rangle_c^2 }
\eeq
where the suffix $c$ stands for connected correlation functions.
The above formulas must be interpreted  according to the aforementioned prescription in the case $n=1$.
From the above expressions we see that the ratio $w_2/w_1$ is precisely equal to the ratio $\omega_2/\omega_1$. We also note that since $w_1$ and $w_2$ are expected to be finite at the transition it is natural to expect that $\omega_1$ and $\omega_2$ diverge as $r^{-3}$ at criticality.
The above expressions can be used to compute the parameter exponent when the Gibbs free energy cannot be directly computed as in the case of fully connected models. In particular we devised a method to compute the above six-point functions for models  defined on finite connectivity random  lattices obtaining a prediction for the ratio $w_1/w_2$ that compares very well with the numerical simulations of the Bethe lattice SG \cite{calta1,PRR}. 
The above expressions can be also used to evaluate the parameter exponent by measuring $\omega_1$ and $\omega_2$ in numerical simulations. In this context it is convenient to consider the fluctuations of the overlaps between different replicas of the same system (additionally they must have the same initial condition in the $n=1$ case). As explained in appendix \ref{fre} one has to consider at least six different replicas in order to evaluate the following eight cubic overlaps: 
\beqa
\W_1 &\equiv & N^2 \langle \delta \Q_{12}\delta \Q_{23}\delta \Q_{31} \rangle 
\\
\W_2  &\equiv &{N^2} \langle \delta \Q_{12}^3  \rangle 
\\
\W_3  &\equiv &{N^2} \langle \delta \Q_{12}^2\delta \Q_{13}  \rangle 
\\
\W_4 & \equiv &{N^2} \langle \delta \Q_{12}^2\delta \Q_{34}  \rangle 
\\
\W_5 & \equiv & {N^2} \langle \delta \Q_{12}\delta \Q_{13}\delta \Q_{24}  \rangle
\\
\W_6 & \equiv &{N^2} \langle \delta \Q_{12}\delta \Q_{13}\delta \Q_{14}  \rangle
\\
\W_7  &\equiv &{N^2} \langle \delta \Q_{12}\delta \Q_{13}\delta \Q_{45}  \rangle 
\\
\W_8 &\equiv &{N^2} \langle \delta \Q_{12}\delta \Q_{34}\delta \Q_{56}  \rangle 
\eeqa
then the six-point cumulants  $\omega_1$ and $\omega_2$ can be obtained using the following formulas \cite{Temesvari02b}: 
\beqa
\omega_1 & = & \W_1 -3 \W_5+3 \W_7-\W_8
\label{omegawprima1}
\\
\omega_2 & = & {1 \over 2}\W_2-3 \W_3+{3 \over 2}\W_4+3 \W_5+2 \W_6 -6 \W_7+2 \W_8 
\label{omegawprima2}
\eeqa
As usual in critical phase transitions we expect cumulants of the order parameter to be divergent. More precisely  we expect that in the case $n=0$ the cumulants $\W$ diverge as $1/r^3$ on the base of the solution of mean-field models. It is to be expected that this feature complicates the numerical evaluation  of $\omega_1$ and $\omega_2$, nevertheless it appears still to be feasible according to preliminary results.

In order to simplify the numerical evaluation one could take advantage of the fact that the eight cumulants defined above are not independent at criticality. This issue is discussed in appendix \ref{critn}, in particular in the case of the transition in a field one can show that at the leading $O(1/r^3)$  order   $\omega_1$ and $\omega_2$ can be expressed in terms of just $\W_1$ and $\W_2$ that can be evaluated considering only three real replicas:
\beqa
\omega_1 & = & {11  \over 30}  \W_1- {2\over 15} \W_2
\\
\omega_2 & = & {4  \over 15}  \W_1- {1\over 15} \W_2
\eeqa
The nature of the relationship between the eight cumulants however depends on the transition and in particular the above relations are only valid for the $n=0$ case provided that only the replicon eigenvalue is critical. This corresponds to the structure of section \ref{contf} but not to that of section \ref{contnof}.
In the case $n=1$, corresponding to MCT, this kind of analysis will be performed in a separate publication, here we only mention that we expect the evaluation of the various cubic cumulants to be more difficult than in the $n=0$ case. Indeed based on a diagrammatic analysis similar to that of \cite{Franz11b} it is expected that all the $\W_i$'s diverge as $1/r^5$ with the same prefactor. Linear combinations can be formed that diverges with powers $1/r^4$ and $1/r^3$. Therefore the $O(1/r^5)$ and $O(1/r^4)$ contributions to $\omega_1$ and $\omega_2$ in (\ref{omegawprima1}) and (\ref{omegawprima2}) should cancel exactly in order to  give  an $O(1/r^3)$ contribution according to eq. (\ref{ow}) possibly leading to large finite size effects.  
Besides these problems the $n=1$ case is also more complicate because of metastability. Indeed states are only defined in the glassy phase away from the dynamical transition, therefore in order to evaluate the various $\W$'s at the critical point one should proceed by extrapolation. Alternatively a more safer procedure is to perform a dynamical evaluation similar to what was done in \cite{Franz11b} for the two point functions $G$. In this case one should sit a the critical point and study six replicas with different thermal histories and the same equilibrium condition evaluating the $\W$'s (and correspondingly $\omega_1$ and $\omega_2$) as a function of time. A parametric plot in power of the average overlap could then be used in order to extract the critical value of the parameter exponent.

We conclude this section noting that the expansion of the Gibbs free energy has actually eight types of cubic terms in $\delta Q_{ab}$. The form of these terms is the same of those of the free energy, see eq. \ref{cubicf} in appendix \ref{fre}. However in eq. (\ref{gibbsfre}) we have displayed only those corresponding to $\sum_{ab}\delta Q_{ab}^3$ and $\Tr \delta Q^3$, this is because they are the only two terms relevant for the present discussion. An explicit computation shows that the remaining terms generate vanishing contributions to the dynamical equation of state both in the continuous and discontinuous case. This is due to the fact that each of these terms turn out to be proportional to differences of one-time quantities computed respectively in the FM limit ($\Gamma_0=\infty$) and at finite $\Gamma_0$. 
Since we are at equilibrium one-time quantities are constant in time and thus independent of the value of $\Gamma_0$ therefore these differences are strictly zero. This is also the reason why in appendix \ref{inverse} we do not report the full Legendre transform inversion but only the expression of $w_1$ and $w_2$.

\section{Conclusions}
\label{conclusions}

We have established a connection between the parameter exponent and the replicated Gibbs free energy. In the case of the fully connected model considered at the beginning  this connection can be explicitly verified. In spite of its simplicity the analysis of this simple mean-field model provides sufficient insight to argue that the result is more general. In order to do so we have  proceeded {\it a la}  Ginzburg-Landau,  {\it  i.e.}  starting from the assumption that, at  some   point in  parameter  space, the given  model displays a phase transition characterized by a  certain  form of the  replicated Gibbs  free energy. In particular we have considered SG transitions that are governed by a RS theory with $n=0$ replicas and dynamical transitions characterized by a RS theory with $n=1$ replicas and the MCT phenomenology. Dynamics is  described  by the  dynamical Gibbs  free  energy that contains  much more  information  than  the  static one. However we have argued that  critical dynamics is governed  by  the  dynamical  Gibbs  free energy in  the  so-called  fast  motion  limit. In this  regime one  sees that the two-point correlation is described  by a scaling function that  obeys  a universal  equation  determined qualitatively by the nature  of the  transition and quantitatively by  a single non-universal parameter (the  parameter  exponent). As we saw in section \ref{SD}  this parameter  is  given  by  the  ratio of two  cubic  vertexes  of the dynamical  Gibbs  free energy in the  fast motion  limit.  The key observation is that in the FM limit the proper  vertexes of the dynamics Gibbs free energy are the same of the static replica theory as expected on the base of standard statistical mechanics.
This last result is particularly evident within the context of the super-field formulation of Langevin dynamics. As we saw in section \ref{SD} going from statics to dynamics is straightforward in this context and the fact that the term $(\delta Q^2)_{ab}$ yields the appropriate memory term in the dynamics can be proved in a few lines (see appendix \ref{appfor}). In a similar way it is straightforward to obtain the critical dynamical equation corresponding to a transition of a given type.

Let us comment that it is rather surprising that the dynamical exponents are completely determined by purely  static quantities. It is also interesting to observe that while the phase transitions considered here are characterized by the divergence of four-point functions, the dynamical exponents are determined by six-point functions. As such there is a precise relationship between the dynamical exponents and the static six-point functions that can be checked directly in simulations and experiments although a major issue is that the six-point functions are also divergent at the critical point. 

Within the context of the replica method eq. (\ref{result}) is of great technical importance. Indeed it was often assumed in the past that the replica method, being essentially a static technique, would only be able to localize the critical temperature and the non-ergodicity parameter but was intrinsically unable to characterize a strictly dynamical quantity like $\lambda$. This is not the case and our result eq. (\ref{result}) has been recently applied in \cite{franz12} to compute $\lambda$ within the Hypernetted-chain approximation of cloned liquid theory. 
On the other hand the universal nature of eq. (\ref{uno}) and of the critical correlators eqs. (\ref{SVDYN1}) and (\ref{SVDYN2}) is not a new result, although one may find it interesting to recover it by means of the replica method. The connection between the $a$ and $b$ exponents was first obtained within schematic MCT theories \cite{Leuthesser84,Bengtzelius84,Gotze84}. Later G{\"o}tze \cite{Gotze85} realized that it is valid under very general assumptions. He considered a generic Mode-Coupling functional  $\hat{F}(f_k)$ and argued that the equations for the critical correlators are universal in the sense that they do not depend on the precise form of $\hat{F}(f_k)$. More precisely they depend on it only through the parameter  $\lambda$ that can be expressed in terms of the second variation of $\hat{F}(f_k)$ at the critical point. 
Note that the derivation within the present paper is also very similar to G{\"o}tze's:  one introduces a rather general functional (the full mode-coupling functional $\hat{F}(f_k)$ in MCT or the replicated Gibbs free energy in the replica approach) and argues that there is universality solely under the assumption that there is a critical point. The parameter $\lambda$ is then expressed in terms of the behavior of the general object near the critical point.   
The exact relationship between $\lambda$ and static quantities represented by eq. (\ref{ris2}) is instead a novel physical prediction that was not obtained previously in the context of MCT. In order to avoid any confusion we recall that in standard MCT one can actually express $\lambda$ in terms of the static structure factor of the liquid but the expression is only approximate. Indeed to compute $\lambda$ within MCT one has to specify the mode-coupling functional $\hat{F}(f_k)$. The standard approximation \cite{Bengtzelius84} is to retain only the two-point vertex $V^{(2)}(q;k,p)$ and then to use Sjogren's  approximation \cite{Sjogren80} that yields $V^{(2)}(q;k,p)$ as a function of the static structure factor of the liquid $S(k)$. As a result the parameter $\lambda$ is expressed in terms of $S(k)$, see {\it e.g.} eq. (18) in \cite{Nauroth97}. Due to the various approximations involved, the result, although often accurate, has not the same status in MCT of eq. (\ref{uno}) and eqs. (\ref{SVDYN1}) and (\ref{SVDYN2}) that were shown to be exact in Ref. \cite{Gotze85}.
The most interesting aspect of eq. (\ref{ris2}) is that the quantities involved can be measured in an experiment or a numerical simulation. Therefore in principle eq. (\ref{ris2}) can be verified measuring independently both $\lambda$ and the cumulants $\omega_2$ and $\omega_1$. For completeness we write down explicitly the formulas of $\omega_1$ and $\omega_2$ for a supercooled liquid: 
\beq
\omega_1 ={1 \over V}\int d\br_1 d\br_2 d\br_3 
\overline{\langle \rho(\br_1)\rho(\br_2)\rangle_c \langle \rho(\br_2)\rho(\br_3)\rangle_c \langle \rho(\br_1)\rho(\br_3)\rangle_c} \ ;
\eeq
\beq
\omega_2= {1 \over V}\int d\br_1 d\br_2 d\br_3 \overline{\langle \rho(\br_1)\rho(\br_2) \rho(\br_3)\rangle_c^2 }\ ;
\eeq
where $\rho(\br)$ is the density. The angle brackets means thermal averages that have to be computed within the same glassy state. The overline means average over the different metastable glassy states. The above quantities can in principle be measured both above or below the critical temperature. As explained in section (\ref{general}) in the latter case the thermal averages should be measured dynamically starting from the same initial condition on the time scale of the $\beta$ regime.

In \cite{Andreanov09} dynamical corrections to standard MCT were computed and it was shown that they lead to a renormalization of the parameter exponent but do not change the form of the critical dynamical equations, in particular the relationship between the exponent $a$ and $b$ remains the same. This result led the authors to suggest that the critical  MCT equation has some universal features that grant to it the status of a Landau theory of the glass transition. Our result shows that this is indeed the case, because the dynamical critical equation follows just from the structure of the replicated Gibbs free energy near the critical point. This means that all models that have mean-field behavior at the static level are described by the MCT critical equation. 
However we stress that the assumption that the MCT transition in models of structural glass can be characterized by a replicated Gibbs free energy of the form considered here is a point that should be put on more solid ground by means of quantitative approximate computations, possibly along the lines of recent encouraging results \cite{Szamel}.
In this respect the present results could be also useful, indeed the computation of the parameter exponent $\lambda$ can be performed in a purely static framework which is usually simpler than the dynamical one. 
     
Below the upper critical dimension one would expect that the ratio $w_2/w_1$ should converge to a universal renormalized value. However we do not expect that this ratio should still coincide with the parameter exponent. Indeed the crucial assumption for the validity of the result is the fact that the Gibbs free energy admits an expansion like those of section \ref{SD} around the critical point. 
This assumption fails below the upper critical dimension and the equation of state is different from those considered here. It is possible that there is still a close connection between the statical and dynamical equation of state but this is a completely open problem that we leave for future work. 

These results can be developed in various directions. One can consider for instance off-equilibrium dynamics in the quasi-static regime. It is well known that aging dynamics below $T_d$ in mean-field SG models is governed by the so-called threshold states that can be characterized through a 1RSB solution with a breaking parameter $x<1$ fixed by the condition that the replicon vanishes \cite{Crisanti93,Cugliandolo93}. In some models \cite{CKL} one observes a $\beta$ regime with two dynamical exponents that obey the following relationship:
\beq
{\Gamma^2(1-a) \over \Gamma(1-2 a)}=x{\Gamma^2(1+b) \over \Gamma(1+2 b)}=\lambda
\label{uno-off}
\eeq
and it would be interesting to check whether $\lambda$ can be computed from an expansion of the Gibbs free energy in the static $1RSB$ solution.
Another interesting case in when the parameter $\lambda$ becomes equal to one. This case has been studied extensively in the structural glass literature \cite{Gotze09} and it would be interesting to understand if the connection between statics and dynamics holds too. Work is under way in this direction and preliminary results shows that it is indeed so and that the quartic coefficients in the Replicated Gibbs free energy becomes relevant, corresponding to eight-point functions. We note that this point corresponds to the end-point of a first order glass transition line in the context of pinning \cite{Cammarota11}.

{\em Acknowledgments.} ~~ We thank F. Caltagirone, U.  Ferrari, S. Franz, L. Leuzzi, F. Ricci-Tersenghi and G. Szamel for useful discussions. The European Research Council has provided financial support through ERC
grant agreement no. 247328.

\appendix

\section{The trace term in the dynamics}
\label{appfor}

In this appendix we compute the boson-boson component of $\int \delta Q(1,3)\delta Q(3,2)  d3$ (that can be obtained setting formally $\theta_1=\theta_2=\overline{\theta}_1=\overline{\theta}_2=0$) and show that it is given by
\beq
\left. \int \delta Q(1,3)\delta Q(3,2)  d3\, \right|_{\theta_1=\theta_2=\overline{\theta}_1=\overline{\theta}_2=0}=-C^2(t)-\int_0^t (C(t-y)-C(t))\dot{C}(y) dy\ .
\label{for1}
\eeq 
where $t=t_1-t_2$. 
Using the definition $\delta Q(1,2)=Q(1,2)-C(0)\delta(1,2)$ we write:
\beq
\int \delta Q(1,3)\delta Q(3,2)  d3 = \int  Q(1,3) Q(3,2)  d3 - 2 C(0)Q(1,2)
\label{for2}
\eeq 
Note that the second term depends on $C(0)$ which is a model-dependent quantity of order $O(1)$. 
Therefore it is disturbing for two reasons: i) it is first order in $Q(1,2)$ and ii) it depends on the non-universal quantity $C(0)$. However we will see that it is canceled exactly by an opposite contribution from the first term.

From now on we set $\theta_1=\theta_2=\overline{\theta}_1=\overline{\theta}_2=0$ and thus we can write:
\beq
Q(1,3)=C(t_1,t_3)+T R(t_1,t_3)\theta_3 \overline{\theta}_3\, , \ \  Q(3,2)=Q(2,3)=C(t_2,t_3)+T R(t_2,t_3)\theta_3 \overline{\theta}_3
\eeq
Where $R(t_1,t_2)$ is the equilibrium response function.
The integration over the variable $3$ selects only some of the terms:
\beq
\int Q(1,3)Q(3,2)d3 = -\int_{-\infty}^{t_1}C(t_2-t_3)\dot{C}(t_1-t_3)dt_3-\int_{-\infty}^{t_2}C(t_1-t_3)\dot{C}(t_2-t_3)dt_3
\eeq
where we have used time translational invariance and the fluctuation-dissipation theorem:
\beq
T R(t)=-\theta (t)\dot{C}(t)
\eeq
The above expression can be split into two parts:
\beq
\int Q(1,3)Q(3,2)d3 = -\int_{t_2}^{t_1} C(t_2-t_3)\dot{C}(t_1-t_3)dt_3-\int_{-\infty}^{t_2}[C(t_1-t_3)\dot{C}(t_2-t_3)+\dot{C}(t_1-t_3)C(t_2-t_3)]dt_3
\eeq
integrating by part and using $C(\infty)=0$ we obtain:
\beq
\int Q(1,3)Q(3,2)d3 = -C^2(t)-\int_0^t (C(t-y)-C(t))\dot{C}(y) dy+2 C(t)C(0)
\eeq
where $t\equiv t_1-t_2$. Substituting the last equation in eq. (\ref{for2}) (evaluated at $\theta_1=\theta_2=\overline{\theta}_1=\overline{\theta}_2=0$) we obtain (\ref{for1}).
An explicit computation shows that the complete expression of $\int \delta Q(1,3)\delta Q(3,2)  d3$ can be obtained by the application of the operator within curly brackets in expression (\ref{genFDT}) to the r.h.s. of eqs. (\ref{for1}).

\section{Free Energy in a field}
\label{fre}
In this appendix we express the coefficients of the free energy of a generic spin-glass model in terms of cumulants of the spin distribution. Indeed the Gibbs free energy is defined as the Legendre transform of the free energy in presence of an appropriate replicated field and therefore in appendix \ref{inverse} we will obtain the relationship between the respective cubic coefficients.
The following discussion applies to the continuous transition case ($n=0$) while at the end we will discuss  the discontinuous case ($n=1$). We proceed in full generality considering  $n$ replicas of a a system of $N$ spins $s_i$ specified by a Hamiltonian $H_J(s)$ depending on some quenched parameter.
For the sake of readability we will repeat some of the definitions already given in the body of the paper. Averages in the replicated system can be rewritten as
\beq
\langle \cdots \rangle \equiv \overline{ \langle \cdots \rangle_J}
\eeq
Where $\langle \cdots \rangle_J$ are thermal averages at fixed couplings $J$ while the overline is the average over the couplings that must be performed reweighting  each disorder realization with the single system partition function to the power $n$: 
\beq
\overline{O_J}=\frac{\int dP(J) O_J Z_J^n}{\int dP(J) Z_J^n}
\eeq
Note that the thermal averages between different replicas factorize prior to the disorder averages.
We define the following free energy functional:
\beq
F(\lambda) \equiv -{1 \over N}\ln \langle e^{\sum_{(ab)}N\lambda_{ab}\delta \Q_{ab}}\rangle
\label{FDEF}
\eeq
where 
\beq
\delta \Q_{ab}={1 \over N}\sum_i s_i^a s_i^b- q
\eeq
and 
\beq
q \equiv {1 \over N}\sum_{i}\langle s_i^a s_i^b \rangle = \sum_i \overline{\langle s_i\rangle_J^2}
\eeq
We note that the above free energy functional arises if we apply to each spin $s_i^a$ of each replica a Gaussian distributed random field $h_i^a$ with covariance matrix given by $\overline{h_i^a h_j^b}=\lambda_{ab}\delta_{ij}$. 
We expand $F(\lambda)$ in powers of $\lambda$ at the third order assuming $\lambda_{aa}=0$ $\forall a$:
\beq
F(\lambda)= -{1 \over 2}\sum_{(ab),(cd)}\lambda_{ab}G_{ab,cd}\lambda_{cd}-{1\over 6} \sum_{(ab),(cd),(ef)}\W_{ab,cd,ef}\lambda_{ab}\lambda_{cd}\lambda_{ef}
\label{FLW}
\eeq
The above expression is fully general, however in a replica symmetric (RS) phase we have only three possible values of $G$ and eight possible values of $W$:
\beq
G_{ab,ab}=G_1\, , \ \ G_{ab,ac}=G_2\, , \ \  G_{ab,cd}=G_3
\eeq
\beq
\W_{ab,bc,ca}=\W_1\,,\ \ \W_{ab,ab,ab}=\W_2\,,\ \ \W_{ab,ab,ac}=\W_3\,,\ \ \W_{ab,ab,cd}=\W_4\,,\ \  
\eeq
\beq
\W_{ab,ac,bd}=\W_5\,,\ \ \W_{ab,ac,ad}=\W_6\,,\ \ \W_{ac,bc,de}=\W_7\,,\ \ \W_{ab,cd,ef}=\W_8\,,\ \ 
\eeq
The cubic part of the free energy can be recast in the following form \cite{Temesvari02b}: 
\beqa
 \sum_{(ab),(cd),(ef)}\W_{ab,cd,ef}\lambda_{ab}\lambda_{cd}\lambda_{ef} & = &  \omega_1 \sum_{abc}\lambda_{ab}\lambda_{bc}\lambda_{ca}+\omega_2 \sum_{ab}\lambda_{ab}^3+        
\nonumber
\\
& +& \omega_3 \sum_{abc}\lambda_{ab}^2\lambda_{ac}+\omega_4 \sum_{abcd}\lambda_{ab}^2\lambda_{cd}+\omega_5 \sum_{abcd}\lambda_{ab}\lambda_{ac}\lambda_{bd}+
\nonumber
\\
& + &  \omega_6 \sum_{abcd}\lambda_{ab}\lambda_{ac}\lambda_{ad}+\omega_7 \sum_{abcde}\lambda_{ac}\lambda_{bc}\lambda_{de}+\omega_8 \sum_{abcdef}\lambda_{ab}\lambda_{cd}\lambda_{ef}                      \, ,
\label{cubicf}
\eeqa
the above identity leads to the following relationships between the $\omega$'s and the $\W$'s \cite{Temesvari02b}:
\beqa
\omega_1 & = & \W_1 -3 \W_5+3 \W_7-\W_8
\\
\omega_2 & = & {1 \over 2}\W_2-3 \W_3+{3 \over 2}\W_4+3 \W_5+2 \W_6 -6 \W_7+2 \W_8 
\\
\omega_3 & = &  3 \W_3 - 3 \W_4 - 6 \W_5 - 3 \W_6 + 15 \W_7 - 6 \W_8
\\
\omega_4 & = & {3 \over 4}(\W_4-2 \W_7+\W_8)
\\
\omega_5 & = & 3\W_5-6\W_7+3\W_8
\\
\omega_6 & = & \W_6-3\W_7+2\W_8
\\
\omega_7 & = & {3 \over 2}\W_7-{3 \over 2}\W_8
\\
\omega_8 & = & {1 \over 8}\W_8
\eeqa
From the definition \ref{FDEF} we easily see that  the coefficients of $F(\lambda)$ can be related to spin averages, in particular $G$ is precisely the dressed propagator:
\beq
G_{(ab),(cd)}\equiv -{\partial^2 \over \partial \lambda_{ab}\partial \lambda_{cd}}F(\lambda)=N\langle \delta \Q_{ab}\delta \Q_{cd}\rangle
\label{DL2}
\eeq
In the following and in the previous expression averages are always computed at $\lambda_{ab}=0$.
Assuming that we are in a RS phase we obtain that $G_{(ab),(cd)}$ can take three possible values depending on whether there are two, three or four different replica indexes. The corresponding values are:
\beqa
G_1 & \equiv & N \langle \delta \Q_{12}^2\rangle = {1 \over N}  \sum_{ij}(\overline{\langle s_i s_j\rangle^2}-q^2) 
\label{G1}
\\
G_2 & \equiv & N \langle \delta \Q_{12} \delta \Q_{13}\rangle = {1 \over N}\sum_{ij}(\overline{\langle s_i s_j\rangle \langle{s_i}\rangle \langle{s_j}\rangle}-q^2) 
\label{G2}
\\
G_3 & \equiv & N \langle \delta \Q_{12} \delta \Q_{34}\rangle = {1 \over N}  \sum_{ij}(\overline{ \langle{s_i}\rangle^2 \langle{s_j}\rangle^2}-q^2)
\label{G3}
\eeqa
The cubic terms are given by the third derivative:
\beq
\W_{(ab),(cd),(ef)}\equiv -{\partial^3 \over \partial \lambda_{ab}\partial \lambda_{cd}\partial \lambda_{ef}}F(\lambda)=N^2\langle \delta \Q_{ab}\delta \Q_{cd}\delta \Q_{ef} \rangle_c =N^2 \langle \delta \Q_{ab}\delta \Q_{cd}\delta \Q_{ef} \rangle 
\label{DL3}
\eeq
where the suffix $c$ stands for connected functions with respect to the overlaps (not with respect to the spins) and the second equality follows from the fact that the average of $\delta Q_{ab}$ is zero by definition.
Depending if some replica indexes are equal the cubic cumulants can take eight possible values:
\beqa
\W_1 & = &N^2 \langle \delta \Q_{12}\delta \Q_{23}\delta \Q_{31} \rangle =
\\
& = & {1 \over N}  \sum_{ijk}\overline{\langle s_i s_j \rangle \langle s_j s_k \rangle \langle s_k s_i \rangle} - {3 q}  \sum_{ij }\overline{\langle s_i s_j\rangle \langle{s_i}\rangle \langle{s_j}\rangle}+ 2 N^2 q^3 
\label{W1} 
\\
\W_2 & = & {N^2} \langle \delta \Q_{12}^3  \rangle =
\nonumber
\\
& = & {1 \over  N}  \sum_{ijk}\overline{\langle s_i s_j s_k\rangle^2 } - {3 q }  \sum_{ij }\overline{\langle s_i s_j\rangle^2}+  2 N^2 q^3 
\label{W2}
\\
\W_3 & = &{N^2} \langle \delta \Q^2_{12}\delta \Q_{13}  \rangle =
\nonumber
\\
& = & {1 \over  N}  \sum_{ijk}\overline{\langle s_i s_j s_k\rangle \langle s_i s_j \rangle \langle s_k \rangle } - {2 q }  \sum_{ij }\overline{\langle s_i s_j\rangle \langle s_i \rangle \langle s_j \rangle}- { q }  \sum_{ij }\overline{\langle s_i s_j\rangle^2}+  2 N^2 q^3 
\label{W3}
\\
\W_4 & = &{N^2} \langle \delta \Q_{12}^2\delta \Q_{34}  \rangle =
\nonumber
\\
& = & {1 \over  N}  \sum_{ijk}\overline{\langle s_i s_j\rangle^2 \langle s_k \rangle^2 } - {2 q }  \sum_{ij }\overline{\langle s_i \rangle^2 \langle s_j \rangle^2 }- { q }  \sum_{ij }\overline{\langle s_i s_j\rangle^2}+  2 N^2 q^3 
\label{W4}
\\
\W_5 & = & {N^2} \langle \delta \Q_{12}\delta \Q_{13}\delta \Q_{24}  \rangle =
\nonumber
\\
& = & {1 \over  N}  \sum_{ijk}\overline{\langle s_i s_j\rangle \langle s_i s_k\rangle \langle s_k \rangle \langle s_j \rangle } - {2 q }  \sum_{ij }\overline{\langle s_i s_j\rangle \langle s_i \rangle \langle s_j \rangle }- { q }  \sum_{ij }\overline{\langle s_i \rangle^2 \langle s_j\rangle^2}+  2 N^2 q^3 
\label{W5}
\\
\W_6 & = & {N^2} \langle \delta \Q_{12}\delta \Q_{13}\delta \Q_{14}  \rangle =
\nonumber
\\
& = & {1 \over  N}  \sum_{ijk}\overline{\langle s_i s_j s_k\rangle \langle s_i\rangle \langle s_j \rangle \langle s_k \rangle } - {3 q }  \sum_{ij }\overline{\langle s_i s_j\rangle \langle s_i \rangle \langle s_j \rangle }+  2 N^2 q^3 
\label{W6}
\\
\W_7 & = & {N^2} \langle \delta \Q_{12}\delta \Q_{13}\delta \Q_{45}  \rangle =
\nonumber
\\
& = & {1 \over  N}  \sum_{ijk}\overline{\langle s_i s_j \rangle \langle s_k\rangle^2 \langle s_i\rangle \langle s_j \rangle } - {2 q }  \sum_{ij }\overline{\langle s_i\rangle^2 \langle s_j\rangle^2 }- {q }  \sum_{ij }\overline{\langle s_i s_j\rangle \langle s_i \rangle \langle s_j \rangle }+  2 N^2 q^3 
\label{W7}
\\
\W_8 & = &{N^2} \langle \delta \Q_{12}\delta \Q_{34}\delta \Q_{56}  \rangle =
\nonumber
\\
& = & {1 \over  N}  \sum_{ijk}\overline{\langle s_i \rangle^2 \langle s_j \rangle^2 \langle s_k\rangle^2 } - {3 q }  \sum_{ij }\overline{\langle s_i\rangle^2 \langle s_j\rangle^2 }+  2 N^2 q^3 
\label{W8}
\eeqa
Substituting the above expressions in the relationship between the $\omega$'s and the $\W$ we obtain:
\beqa
\omega_1 & = & {1 \over N} \sum_{ijk}\overline{\langle s_i s_j\rangle_c \langle s_j s_k\rangle_c \langle s_k s_i\rangle_c}
\label{omega1}
\\
\omega_2 & = & {1 \over 2 N} \sum_{ijk}\overline{\langle s_i s_j s_k\rangle_c^2 }
\label{omega2}
\\
\omega_3 & = &{3 \over  N}  \sum_{ijk}\overline{\langle s_i s_j s_k\rangle_c \langle s_i s_j \rangle_c \langle s_k \rangle }
\label{omega3}
\\
\omega_4 & = & {3 \over 4 N}  \sum_{ijk}\left[\overline{\langle s_i s_j\rangle_c^2 \langle s_k \rangle^2 }-\overline{\langle s_i s_j\rangle_c^2}\ \overline{ \langle s_k \rangle^2 }\right]
\label{omega4}
\\
\omega_5 & = & {3 \over  N}  \sum_{ijk}\overline{\langle s_i s_j\rangle_c \langle s_i s_k\rangle_c \langle s_k \rangle \langle s_j \rangle }
\label{omega5}
\\
\omega_6 & =& {1 \over  N}  \sum_{ijk}\overline{\langle s_i s_j s_k\rangle_c \langle s_i\rangle \langle s_j \rangle \langle s_k \rangle }
\label{omega6}
\\
\omega_7 & = & {3 \over 2 N}\sum_{ijk}\left[\overline{\langle s_i s_j \rangle_c\langle s_i\rangle \langle s_j\rangle \langle s_k\rangle^2}-\overline{\langle s_i s_j \rangle_c\langle s_i\rangle \langle s_j\rangle}\ \overline{\langle s_k\rangle^2}\right]
\label{omega7}
\\
\omega_8 & = & {N^2 \over 8}\overline{\left( q_J-q\right)^3}
\label{omega8}
\eeqa
We remark that the above expression show that upon passing from the $\W$'s to the $\omega$'s there is increase in symmetry, in particular we see that due to various cancellations $\omega_1$,  $\omega_2$,  $\omega_3$,  $\omega_5$ and $\omega_6$ have a single disorder average, $\omega_4$ and $\omega_7$ have two disorder average and only $\omega_8$ has three disorder averages. 

The above discussion is valid in the case $n=0$, in the case of a discontinuous transition the final expressions are the same but have to be interpreted in a different way.
As we have already said in section \ref{general} below the dynamical transition temperature there is an exponential number of metastable states. Correspondingly {\it the thermal averages in the above expressions have to be performed within the same metastable state, while the overline stands for both the summation over all possible states and the standard disorder average}.
In the following we discuss more extensively this issue and also the fact that the relevant Gibbs free energy of the problem is RS with $n=1$ replicas. After all the starting point in the replica framework is the same of the continuous transition, that is the computation of the free energy of a  system replicated $n$ times with $n \rightarrow 0$ and one may wonder why we end up considering instead  $n \rightarrow 1$.

One way to characterize the presence of many metastable states below dynamical transition is to select an equilibrium configuration of the system (the reference configuration) and then study the effect of a small field pointing in the direction of the reference configuration \cite{Franz95a}. When the field goes to zero the overlap between a configuration of the constrained system and the reference configuration should be equal to the equilibrium value $q_0$ of the overlap between two independent copies. Instead below the dynamical transition temperature the overlap with the reference configuration goes to a higher value $q_1$ because the system remains stuck in the metastable state to which the reference configuration belongs.
The smaller value $q_0$ corresponds to the overlap between configurations in different states, while $q_1$ corresponds to the overlap between configurations in the same state. However the overlap between two independent replicas remains $q_0$ because the probability of extracting two configurations from the same state vanishes since there is an exponential number of states with essentially equal (and therefore infinitesimal) probability.

It is well known \cite{Monasson95,Franz95a,Mezard1999} that in the replica framework this feature is encoded by  spontaneous replica-symmetry breaking. More precisely above the transition the equation of state of the order parameter (the $n\times n$ overlap matrix $q_{ab}$) has only the solution $q_{ab}=q_0$ while below a new solution appears. This solution is characterized by the fact that there is a subgroup of size $m$ of the $n$ replicas such that the overlap is $q_1$ i.f.f. both replicas are inside the subgroup while is $q_0$ otherwise. Analytical continuation has to be taken with $m\rightarrow 1$ and $n \rightarrow 0$. 

As we said in section \ref{SD} in order to study critical dynamics we may consider $n'\rightarrow 0$ static replicas of the system in order to select the (equilibrium) reference configuration and then we may consider its dynamical evolution.
In this mixed dynamical-replica framework the system of the initial configuration and of its evolution at later times corresponds in the FM limit to a set of $m=1+0$ replicas. The key point is that the  overlap between the initial configuration and the remaining $n'-1$ replicas is equal to $q_0$ and does not change considering the time evolution of the initial equilibrium configuration therefore at all orders in the expansion around the FM limit we can set $\delta Q(a,b)=0$ whenever one of the two indexes corresponds to one of the $n'-1$ replicas. This means that the in order to study the correction to the FM limit in the dynamics we only need to take into account deviations of the order parameter inside the block with $q_{ab}=q_1$ and this leads to a replica symmetric Gibbs free energy with $n=1$. 
Going back to eq. (\ref{FDEF}) we must consider non-zero values of $\lambda_{ab}$ only inside a $m \times m$ block with $m=1$. As a consequence the replica indexes in eq. (\ref{DL2}) and (\ref{DL3}) are all from  the  same block and therefore the thermal averages in the r.h.s. of eqs. (\ref{G1},\ref{G2},\ref{G3}) and eqs. (\ref{W1},\ref{W2},\ref{W3},\ref{W4},\ref{W5},\ref{W6},\ref{W7},\ref{W8}) have to be computed with different thermal histories but within the {\it same} metastable state. More explicitly in the case of the discontinuous transition we should rewrite eqs. (\ref{omega1}) and (\ref{omega2}) as: 
\beqa
\omega_1 & = & {1 \over N} \sum_{ijk}\overline{\sum_{\alpha}P_{\alpha}\langle s_i s_j\rangle_c^{\alpha} \langle s_j s_k\rangle_c^{\alpha} \langle s_k s_i\rangle_c^{\alpha}}
\label{omega1dis}
\\
\omega_2 & = & {1 \over 2 N} \sum_{ijk}\overline{\sum_{\alpha}P_\alpha(\langle s_i s_j s_k\rangle_c^{\alpha})^2 }
\label{omega2dis}
\eeqa
where the overline is the usual disorder average, $\alpha$ labels the different metastable states, $\langle \dots \rangle^{\alpha}$ means thermal average inside metastable state $\alpha$ and $P_{\alpha}$ is the thermodynamic weight of the state $\alpha$. This expression holds also in the case of model with self-induced disorder where there is no overline but only the summation over different components.  

We note that in \cite{ferra1,calta2,calta3} the discontinuous transition in various fully connected models were studied. These models have the peculiar property that $q_0=0$, as a consequence it is almost immediate  to see the connection of the problem with a RS free energy with $n=1$. In order to avoid a misinterpretation of these results we stress that according to the above discussion the connection is more general and does not at all require $q_0 = 0$.

\section{The inversion of the Legendre Transform}
\label{inverse}

The Gibbs Free energy is defined as the Legendre transform of the Free energy $F(\lambda)$:
\beq
G(\delta Q) \equiv F(\lambda)+\sum_{(ab)}\lambda_{ab}\delta Q_{ab}
\eeq
where $\lambda$ is a function of $\delta Q_{ab}$ according to the following implicit equation:
\beq
\delta Q_{ab}=-{\partial F \over \partial \lambda_{ab}}
\eeq
On the other hand the free energy is the Legendre transform of the Gibbs free energy and we have:
\beq
\lambda_{ab}={\partial G \over \partial \delta Q_{ab}} \ .
\eeq
We consider the expansion of the free energy at third order taking into account only the term $\Tr \lambda^3$ and $\sum_{ab} \lambda^3$ because the other terms are not relevant in the present context, see  the end of section \ref{general}:
\beq
F(\lambda)= -{1 \over 2}\sum_{(ab),(cd)}\lambda_{ab}G_{ab,cd}\lambda_{cd}-{\omega_1 \over 6} \Tr \lambda^3-{\omega_2 \over 6} \sum_{ab} \lambda_{ab}^3
\eeq
Differentiating the free energy with respect to $\lambda_{ab}$ we obtain (using Einstein's sum convention):
\beq
\delta Q_{ab}= G_{(ab),(cd)}\lambda_{cd}+\omega_1 (\lambda^2)_{ab}+\omega_2 \lambda_{ab}^2
\eeq 
From this we obtain
\beq
\lambda_{ab}=M_{(ab),(cd)}\delta Q_{cd}- \omega_1 M_{(ab),(cd)}(\lambda^2)_{cd}- \omega_2 M_{(ab),(cd)}\lambda^2_{cd}
\label{MMM}
\eeq
Where $M$ is the inverse propagator that can be rewritten in the form :
\beq
M_{ab,cd}=M_3+r (\delta_{ac}\delta_{bd}+\delta_{ad}\delta_{bc})+(M_2-M_3) (\delta_{ac}+\delta_{ad}+\delta_{bc}+\delta_{bd})
\eeq
and $r \equiv (G_1-2 G_2+G_3)^{-1}=M_1-2 M_2+M_3$ is the replicon eigenvalue.
At linear order we have:
\beq
\lambda_{ab}=r \delta Q_{ab}+ (M_2-M_3)\sum_c (\delta Q_{ac}+\delta Q_{bc})+M_3 \sum_{(cd)}\delta Q_{cd}
\eeq 
and we can substitute the above expression in (\ref{MMM}) in order to determine $\lambda$ at second order in $\delta Q$. Since in the end we are only interested in terms the form $(\delta Q)_{ab}^2$ and $\delta Q_{ab}^2$ in which both indexes $a$ and $b$ appear we can replace:
\beq
\omega_1 M_{(ab),(cd)}(\lambda^2)_{cd} \longrightarrow r^3 \omega_1 (\delta Q)^2_{ab}
\eeq
\beq
\omega_2 M_{(ab),(cd)}\lambda^2_{cd} \longrightarrow r^3 \omega_2 \delta Q^2_{ab}
\eeq
We define the coefficients of the expansion of the Gibbs free energy according to:
\beq
G(\delta Q)={1 \over 2}\sum_{(ab),(cd)}\delta Q_{ab} M_{ab,cd} \delta Q_{cd}-{w_1 \over 6}\Tr \delta Q^3-{w_2 \over 6}\sum_{ab}\delta Q_{ab}^3 \ .
\eeq
With the above definition we obtain:
\beq
w_1=r^3 \omega_1 \ \ \ w_2=r^3 \omega_2
\eeq

\section{Critical Behavior of the Cubic Cumulants}
\label{critn}

The cubic cumulants $\W_i$ defined previously are all divergent at criticality.
Nevertheless depending on the nature of the transition one can obtain some relationship between their singular parts.
We will consider the case of a $n=0$ RS critical point where the replicon vanishes corresponding {\it e.g.} to the dAT line leaving the study of the $n=1$ RS critical point where the replicon and longitudinal eigenvalues vanish to a separate publication.

We start recalling the relationships that hold at criticality between the quadratic cumulants $G_1$, $G_2$ and $G_3$ in the $n=0$ case:
\beq
G_1-2 G_2=0 \, , \ \ G_1-3 G_3=0 
\eeq
In the above relationship it is intended that we are considering the critical part, indeed each $G_i$ is diverging as $1/r$ approaching the critical point but the above differences remain finite, {\it i.e.} their singular part is zero.
The above relationships can be obtained in various ways. For instance we can use that following relationship:
\beq
\sum_a\langle O \delta Q_{ab}\rangle=0
\eeq  
where $O$ is any observable. It follows from the fact that summing over a replica index we are removing the singular replicon component from $\delta Q_{ab}$ and we are left with only the regular anomalous and longitudinal component. 

Another possible way of obtaining the above result is by considering the expression of the longitudinal and anomalous eigenvalues in terms of $G_1$, $G_2$ and $G_3$ (see \cite{Temesvari02b}) and impose that, contrary to the replicon, they remains finite, {\it i.e.} the singular part of their  inverse is zero. However in this case one must be careful in taking the limit $n \rightarrow 0$ because the longitudinal and anomalous eigenvalue are degenerate in this limit and one remains with a single equation. In order to obtain the second equation one must consider also their difference divided by $n$ and impose that it remains finite in the limit $n \rightarrow 0$.

The above procedures can be used to obtain similar relationship between the cubic cumulants\footnote{ Concerning the second procedure we note that from formulas (48) in \cite{Temesvari02b} since only the replicon eigenvalue is critical we have $g_3=g_4=g_5=g_6=g_7=g_8=0$ that can be translated into six equations for the $\W_i$'s using eqs. (49) in \cite{Temesvari02b} and the eqs. below (21) connecting the $\omega_i$'s and the $\W_i$'s. However again because of the degeneracy between the longitudinal and anomalous eigenvalues it is convenient to consider the following combinations  $g_3=0$, $(g_3-g_4)/n=0$, $g_6=0$, $(g_7-g_6)/n=0$, $(g_7 - g_6 - 1/3 (g_8 - g_6))/n^2=0$ which remains independent in the $n \rightarrow 0$ limit}.
The resulting system of equations is:
\beqa
2 \W_3  & = & \W_2
\\
3\W_4 & = &  \W_2
\\ 
6 \W_5  &=&  -\W_2 + 15 \W_8
\\
 3 \W_6 &=& \W_2 
\\
4 \W_7 &=& 5 \W_8
\\
2 \W_1 &=& -2 \W_2 + 15 \W_8
\eeqa
In principle the expression of the $\omega$'s in terms of the $\W_i$  requires all the eight $\W_i$ (see the previous section), instead the above formulas can be used to obtain an expression the depend just on two of them. For instance if we express everything in terms of three replica cumulants ({\it i.e.} $\W_1$ and $\W_2$) we have:
\beqa
\omega_1 & = & {11  \over 30}  \W_1- {2\over 15} \W_2
\\
\omega_2 & = & {4  \over 15}  \W_1- {1\over 15} \W_2
\eeqa
Note that these relationship holds on the dAT line but cannot hold on its zero-field end point. Indeed in zero field we must have $\W_2=\omega_2=0$ while $\W_1$ remains finite. This apparent contradiction is solved noticing that at the end point also the longitudinal and anomalous eigenvalue vanish producing divergent corrections to the above formulas.

\end{document}